# Towards a new metamodel for the Task Flow Model of the Discovery Method

Carlos Alberto Fernández-y-Fernández
*Instituto de Computación*
*Universidad Tecnológica de la Mixteca*
*Huajuapan de León, Oaxaca, **México***
caff@mixteco.utm.mx

**Abstract**

*This paper presents our proposal for the evolution of the metamodel for the Task Algebra in the Task Flow model for the Discovery Method. The original Task Algebra is based on simple and compound tasks structured using operators such as sequence, selection, and parallel composition. Recursion and encapsulation were also considered. We propose additional characteristics to improve the capabilities of the metamodel to represent accurately the Task Flow Model.*

## 1 Introduction

Visual modelling is the modelling of a computer program or larger software system using graphical notations to develop a model, expressed as one or more diagrams. The model is intended to capture the essentials parts of a system [1] and is used to represent the business processes from a user-centred, or stakeholder's perspective. It contributes to the understanding of the business domain and helps later in the design of the information system. The Unified Modeling Language (UML) is at present the standard visual modelling notation. At the time of writing, it provides thirteen different diagrams that can be used to represent a software system from different aspects and perspectives [2], [3]. The Unified Modeling Language (UML) [2] is an eclectic set of notations for modelling object-oriented designs. Under the supervision of the Object Management Group (OMG) since 1997, the notation set has grown larger and complex [3], to accommodate the concerns of different stakeholders in business and industry. This has led to some criticisms regarding the open-ended semantics and the lack of direction given in modelling [4], [5].

Although there are some accounts of their use in industry (basically in critical systems), the majority of software houses in the "real world" have preferred to use visual modelling as a kind of "semi-formal" representation of software [6].

### 1.1 The Discovery Method

The Discovery Method is an object-oriented methodology proposed formally in 1998 by Simons [7], [8]; it is considered by the author to be a method focused mostly on the technical process. The Discovery Method is organised into four phases; Business Modelling, Object Modelling, System Modelling, and Software Modelling (Simons, pers. Comm.). The Business Modelling phase is task-oriented. A task is defined in the Discovery Method as something that "has the specific sense of an activity carried out by stakeholders that has a business purpose" (Simons, pers. comm.). This task-based exploration will lead eventually towards the two kinds of Task Diagrams: The Task Structure and Task Flow Diagrams. The workflow is represented in the Discovery Method using the Task Flow Diagram. It depicts the order in which the tasks are carried out in the business, expressing also the logical dependency between tasks. While the notation used in the Dis- covery Method is largely based on the Activity Diagram of UML, it maintains consistently the labelled ellipse notations for tasks.

### 1.2 Process Algebra

The term process algebra or process calculus is used to define an axiomatic approach for processes. There is not a unique definition for processes although Baeten [9] says a process refers to the behaviour of a system. Process Algebras have been used to model concurrent systems [10]. Common concepts in the different process algebras are process (sometimes called agent) and action [11]. A process can be seen as any concurrent system with a behaviour based in discrete actions. An action is considered something that happens instantaneously and it is atomic. An action is expressed in conjunction with other actions, using particular operations defined by the algebras.

Some of the principal process algebras comprise ACP, CCS, CSP, and more recently Pi-Calculus. The term process algebra was coined by Bergstra and Klop in the paper [12] where the Algebra of Communicating Processes (ACP) was presented. The Calculus of Communicating Systems (CCS) was proposed by R. Milner [13]. The contrasting calculus of Communicating Sequential Processes (CSP) was proposed by Hoare [14]. An extension and revision to CCS, the Pi-calculus was later proposed by Milner [15].

### 1.3 Why yet Another Process Algebra?

We proposed a simple process algebra that is a direct representation of the Task Flow model for the Discovery Method. Translations from the diagrams to the Task Algebra could easily be made by software developers, which are usually not experienced in formal methods, or automatically generated by an Integrated Development Environment (IDE). Task Flow models are tipically used in the Businness Modelling phase of the Discovery Method or can be used to substitute the Activity diagram of UML [16]. Our Task Algebra attempts to be a lightweight formal method that allows software developers to reduce the effort of writing and verifying formal specifications.

## 2 An overview of the Task Model Project

### 2.1 The Task Algebra for Task Flow models

In our original proposal, the basic elements of the abstract syntax are: the simple task, which is defined using a unique name to distinguish it from others; ε representing the empty activity; and the success σ and failure φ symbols, representing a finished activity. Simple and compound tasks are combined using the operators that build up the structures allowed in the Task Flow Model. The basic syntax structures for the Task Flow Model are sequential composition, selection, parallel composition, repetition, and encapsulation. The algebra definition is shown in Table 1.

*Table 1: Abstract syntax definition*

```
Activity ::= ε              -- empty activity
        | σ                 -- succeed
        | φ                 -- fail
        | Task              -- a single task
        | Activity ; Activity   -- a sequence of activity
        | Activity + Activity   -- a selection of activity
        | Activity || Activity  -- parallel activity
        | μx.(Activity ; ε + x)   -- until-loop activity
        | μx.( ε + Activity ; x)  -- while-loop activity

Task::= Simple              -- a simple task
      | { Activity }        -- encapsulated activity
```

A task can be either a simple or a compound task. Compound tasks are defined between brackets '{' and '}', and this is also called encapsulation because it introduces a different context for the execution of the structure inside it. Defining a compound task takes the form of:

(1) `let Task={Activity}`

Curly brackets are used in the syntax context to represent diagrams and sub-diagrams but these also have implications for the semantics. Also, parentheses can be used to help comprehension or to change the associativity of the expressions. Expressions associate to the right by default. Further information of our metamodel and the axioms can be found in [17], [18].

Subsequently, the precise semantics for the abstract task algebra was developed. The semantics was designed in terms of trace sets representing all possible complete execution paths for a system of tasks. The soundness and congruence of the task algebra was proved in [17]. In addition, an implementation of the algebra in the Haskell programming language was developed. The traces generated by the program were then analysed by submitting these to queries about (in-)equality, tested using set operations, and more general theorems about temporal logic properties, tested using LTL and CTL theorems.

### 2.2 Software supporting our metamodel

Additionally, a graphical tool was needed to generate the diagrams and to translate them into the algebra. This tool supports the automatic construction and simplification of formal models by developers, directly from diagram specifications. Also, further work was invested in visualising the results from applying LTL and, particularly, CTL queries, which are not easy to interpret in their current form. A graphical browser, capable of visualising and navigating over trees of traces helps the developer to understand the results of queries. As mentioned above, all these were already available as console tools developed in Haskell to prove the feasibility of the proposal but, in order to be used by real-world developers, an IDE encapsulating the console software was necessary.

Being Eclipse one of the most used environments for software development, we built a Eclipse-based tool that allows modelling and testing of software models that are defined usually in the specification phases [19]. Figure 1 shows the architecture of the Task Model Tool developed. With these tools, developers are not required to increase the quantity of artifacts when developing software. If developers create Task Flow diagrams, they will have a formal specification for their software which could improve communications using the unambiguous notation. In addition, using model-checking software in early stages, may increase the confidence that goes from a correct definition to the final design.

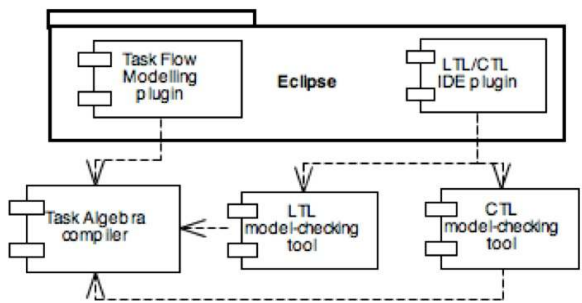

*Figure 1. Architecture of the Task Model Tool.*

The plug-ins developed facilitate the formal specification of the Task Flow models and the verification of these models in a visual and simple way. The queries are structured visually and with it the interpretation of results is even simpler. Every module is easy to use and to understand for programmers due to its integration with Eclipse. Figure 2 shows a view of the user interface of the Eclipse-based development for the Task Flow modelling.

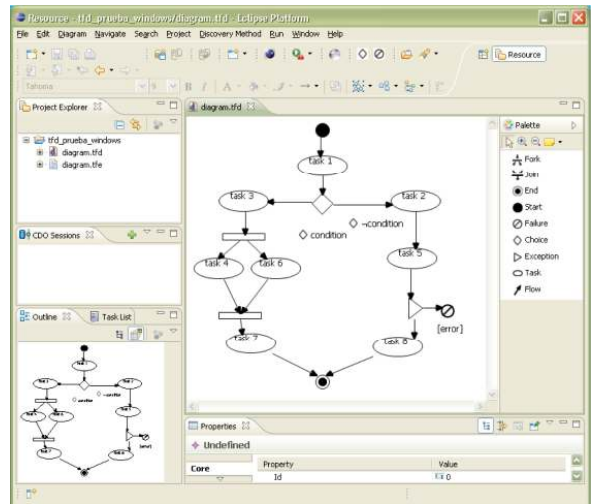

*Figure 2. Partial view of the Eclipse-based tool for Task Flow Modelling.*

## 3 What is next?

Although the work on formalisation of the task algebra is finished, some of the software tools are just reaching their final development. However, we are already looking for improvement in some aspects such as:

- the metamodel,
- the semantics, and
- the software (visual and command line)

### 3.1 Methodology

In Figure 3, we show how we will be developing our research. We will be proposing the change to the syntax of the metamodel. These changes will need to be supported by adaptations of the axioms and semantics already defined. For the semantics, a revision of the trace domain, mapping functions and semantic functions will be needed. In addition, we will have to prove again properties such as soundness and congruence. Completeness was left as future work in the original version of the task algebra but it is a desirable property of any formal system. Finally, all these changes will have an impact in the current set of tools (command line and eclipse-based plugins mainly).

Some work will have to be done adapting or migrating these software tools.

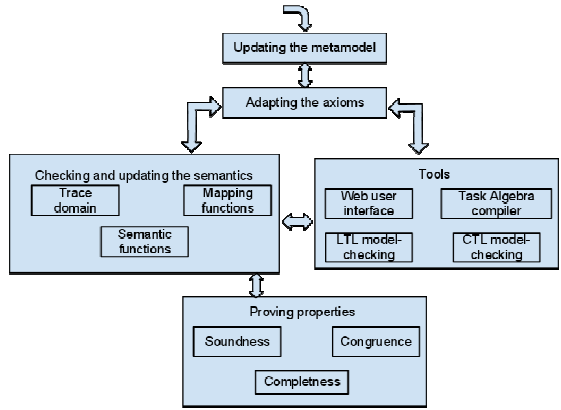

*Figure 3. Descriptive diagram of the methodology for our research.*

### 3.2  Improving the metamodel

For the new version of our algebra, we are proposing basically two strong improvements to the metamodel:

- Guards
- Simple task postcondition

As mentioned in section 2, the original version of the metamodel supports most of the properties available in the Task Flow diagram. The only missing property is the representation of guards. We are proposing to add support for mapping the guards from the task diagrams into the algebra. The guards should allow the use of values, variables and logical operators, which are allowed by the Task Flow diagrams in the Discovery Method. Modelling guards will have clearly an impact in the semantics and it will enhance the kind of queries to be applied over the models. So far, we think guards should be attached next to the selection operator:

```
(2) Activity[guard]+[guard]Activity
```

Where the specific syntax for the guard is still being defined but should allow the use of relational and boolean operators over tasks.

Additionally, as defined originally in [17], [18], task algebra expressions are associative over the right, therefore we are showing examples of equivalent expressions with parenthesis for two and three two selection operators:

```
(3) x[guard]+[guard]y[guard]+[guard]z
    =x[guard]+[guard](
y[guard]+[guard]z )

(4) w[guard]+[guard]x[guard]+
[guard]y[guard]+[guard]z
    = w[guard]+[guard](
x[guard]+[guard]( y[guard]+[guard]z))
```

As mentioned above, we are also proposing the possibility of adding data to simple tasks. With the current version of the task algebra, it is possible to establish whether a simple task is happening at some point but no more information is provided. Additional information could be added to a simple task in as depicted in (5).

```
(5) let task=[list of properties]
```

Where the list of properties will provide extra information for the simple task. Because of this optional extra information, simple tasks are not going to be always atomic. As a consequence, the software designer should be able to analyse beyond the name of the task. These properties could remain or change their values if the task is used more than once in a task flow diagram (6).

```
(6)simpleTask(changing properties)
```

So far, we are not considering compound tasks, because they are already formed by simple tasks. Figure 4 depicts the Task Flow diagram for the process of logging to a system.

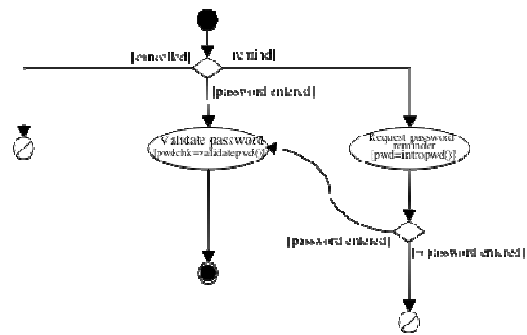

*Figure 4. Login Task Flow diagram.*

The login Task Flow diagram can be translated using the metamodel syntax proposed. The corresponding expression showing task, task properties and conditions should be as follows:

```
{ ( φ [cancelled] +
(validatePassword(pwdchk=validatepwd(
)) [password entered] + [remind] (
requestPassword(pwd=intropwd());
(validatePassword(pwdchk=validatepwd(
) ) [password entered] + [¬ password
entered] φ ) )))}
```

### 3.3 Adapting the semantics

It is easy to see that the changes proposed to the metamodel imply that the semantics will have to be updated as well. The formal semantics developed in [17] is sufficient to demonstrate how the representation of simple and compound tasks is unified at the lowest level of representation. This is enough to represent the tasks in some detail, to determine whether or not these tasks succeed or fail in their execution. Nevertheless, the semantics abstracts over the details of choices taken in conditional expressions and does not further analyse simple tasks, which are considered atomic. The proposed changes to the metamodel have to include an adaptation in the semantics. The semantics should consider including a suitable abstract representation of the states tested in conditional expressions. This could be developed, if simple tasks were decomposed further to express, in some form, how they affected system states, which could be modelled in the semantics as the atomic postconditions of each task. This would support more detailed reasoning about the triggering of different branches and the ability to handle exceptional cases, based on satisfied and unsatisfied atomic preconditions.

### 3.4 Proving properties

Adaptation for the metamodel and the semantics need to be formally represented. We will have to prove again properties such as soundness and congruence for the new algebra. Completness was left as future work in the original version of the task algebra but it is a desirable property of any formal system.

### 3.5 Updating the tools

Tools have to be adapted to the new metamodel and its semantics. Eclipse-based plugins are already working for the original version and should be modified.

Implementations for the Task Algebra, LTL and CTL model-checking tools will have to be adapted or migrated to the new Task Algebra definition.

Additionally, we are currently developing a set of additional software implementations based on the Task Algebra metamodel, such as:

- A Web IDE taking advantage of HTML5 to offer a modern and multi-platform user interface for Task Model development. Figure 5 shows a low-fidelity prototype programmed by a couple of students working temporarily in the project[1].
- A Java API for the task algebra compiler and the LTL/CTL model-checking tools. The original software was developed in Haskell and we are looking for better ways to integrate this software with Eclipse and the Web IDE.
- Using the Kinect technology in the Web IDE, experimenting with the user interface design. We are exploring new paradigms and aim to create an alphabet for human computer interaction when creating diagrams.

Logically, these developments will have to be built and adapted while developing our research.

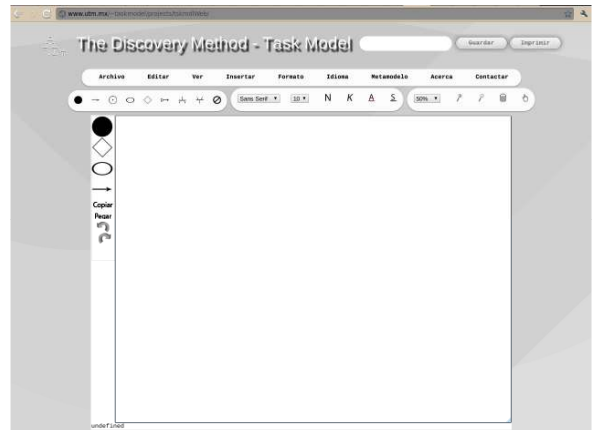

*Figure 5. User interface for web (Low-fidelity prototype).*

## 4 Conclusions

We presented our proposal for the evolution of the Task Algebra in the Task Flow model for the Discovery Method. The original Task Algebra was based on simple and compound tasks structured using operators such as sequence, selection, and parallel composition. Recursion and encapsulation were also considered.

---

[1] This early prototype was created by a couple of students working in their professional practices in our university.

We proposed to add support for mapping the guards from the task diagrams into the algebra. The guards should allow the use of values, variables and logical operators, which are allowed by the Task Flow diagrams in the Discovery Method. Modelling guards will have clearly an impact in the semantics and it will enhance the kind of queries to be applied over the models. We also proposed having simple tasks postconditions, optionally adding data to simple tasks. Decomposing simple tasks to express, in some form, how they impact system states. Logically, our proposal will affect the semantics, proved properties and tools already created.

These apparently small changes are important and will allow us to create much richer Task Flow models, increasing the capability to validate these models in a lightweight formal language.

Finally, we think this research is relevant mainly for three reasons: first, even when the task flow models were defined by Simons originally for the business modelling phase of the Discovery Method, this diagram could also be integrated into other areas of modelling. Like the activity diagram from UML, task flow diagrams could be used during the requirements phase to represent the flow of events of use cases; during analysis and design phases, they could be used to help define the behaviour of operations. Secondly, using a lightweight formal method and model-checking techniques to verify the model could help to decrease the uncertainty along the software development phases. Lastly, it is a common opinion that software engineering is not going to be considered a real engineering discipline until, among other things, we can use tools and techniques to create mathematical models where we can analyse and test solutions. This is where we stand, we are making an small approximation to the formalization of the software engineering discipline.

## 5  Acknowledgment

This work has been funded by the Universidad Tecnológica de la Mixteca.